\documentclass[aps,prd,reprint,amssymb,amsmath,%
  preprintnumbers,showpacs,showkeys
  ]{revtex4-1}
\pdfoutput=1
\usepackage[pdftex,breaklinks,colorlinks=true,allcolors=blue]{hyperref}

\newcommand{\rd}{\mathrm{d}}
\newcommand{\MBint}[1]{\int_{-i\infty}^{i\infty}\frac{\rd{#1}}{2i\pi}}
\DeclareMathOperator{\Rep}{Re}
\DeclareMathOperator{\Imp}{Im}
\newcommand{\hypF}[2]{{{}_{#1}F_{#2}}}

\begin{document}

\preprint{TTK-12-46, SFB/CPP-12-84}

\title{New proofs for the two Barnes lemmas and an additional lemma}

\date{November 12, 2012}

\author{Bernd Jantzen}
\email{jantzen@physik.rwth-aachen.de}
\affiliation{Institut f\"ur Theoretische Teilchenphysik und Kosmologie,
  RWTH Aachen University, 52056~Aachen, Germany}

\begin{abstract}
  Mellin--Barnes (MB) representations have become a widely used tool for
  the evaluation of Feynman loop integrals appearing in perturbative
  calculations of quantum field theory. Some of the MB integrals may be
  solved analytically in closed form with the help of the two Barnes lemmas
  which have been known in mathematics already for one century.
  The original proofs of these lemmas solve the integrals by taking
  infinite series of residues and summing these up via hypergeometric
  functions.
  This paper presents new, elegant proofs for the Barnes lemmas which only
  rely on the well-known basic identity of MB representations, avoiding any
  series summations. They are particularly useful for presenting and
  proving the Barnes lemmas to students of quantum field theory without
  requiring knowledge on hypergeometric functions.
  The paper also introduces and proves an additional lemma for a MB
  integral $\int\!\rd z$ involving a phase factor $\exp(\pm i\pi z)$.
\end{abstract}



\maketitle

\section{\label{sec:intro}Introduction}

When working with generalized hypergeometric functions and relating them
through complex contour integrals, about one century ago, Barnes introduced
the two following identities:
The \emph{first Barnes lemma}~\cite{Barnes:1908},
\begin{align}
  \label{eq:Barnes1}
  &\MBint{z} \, \Gamma(\alpha_1-z) \, \Gamma(\alpha_2-z) \,
    \Gamma(\beta_1+z) \, \Gamma(\beta_2+z)
\nonumber \\*
  &\quad{} = \frac{\Gamma(\alpha_1+\beta_1) \, \Gamma(\alpha_1+\beta_2) \,
    \Gamma(\alpha_2+\beta_1) \, \Gamma(\alpha_2+\beta_2)}{
    \Gamma(\alpha_1+\alpha_2+\beta_1+\beta_2)}
  \,,
\end{align}
and the \emph{second Barnes lemma}~\cite{Barnes:1910},
\begin{align}
  \label{eq:Barnes2}
  &\MBint{z} \, \frac{\Gamma(\alpha_1-z) \, \Gamma(\alpha_2-z) \,
      \Gamma(\beta_1+z) \, \Gamma(\beta_2+z)}{
    \Gamma(\alpha_1+\alpha_2+\beta_1+\beta_2+\beta_3+z)}
\nonumber \\*
  &\qquad{}\times
    \Gamma(\beta_3+z)
\nonumber \\
  &\quad{} = \frac{\Gamma(\alpha_1+\beta_1) \, \Gamma(\alpha_1+\beta_2) \,
      \Gamma(\alpha_1+\beta_3)}{
    \Gamma(\alpha_1+\alpha_2+\beta_1+\beta_2) \,
      \Gamma(\alpha_1+\alpha_2+\beta_1+\beta_3)}
\nonumber \\*
  &\qquad{}\times
    \frac{\Gamma(\alpha_2+\beta_1) \, \Gamma(\alpha_2+\beta_2) \,
      \Gamma(\alpha_2+\beta_3)}{
    \Gamma(\alpha_1+\alpha_2+\beta_2+\beta_3)}
  \,.
\end{align}
These identities involve so-called Barnes contour integrals whose paths of
integration run parallel to the imaginary axis and are curved where
necessary to ensure that they separate positive and negative sequences of
poles in the integrands. Explicitly, the \emph{right poles} of gamma
functions of the type $\Gamma(\alpha_i-z)$ (at
$z=\alpha_i,\alpha_i+1,\ldots$) must lie to the right of the integration
contour, whereas the \emph{left poles} of gamma functions
$\Gamma(\beta_j+z)$ (at $z=-\beta_j,-\beta_j-1,\ldots$) lie to the left of
it.
The two Barnes lemmas are valid whenever such an integration contour can
be found, i.e.\ unless any of the $\alpha_i+\beta_j$ is zero or a negative
integer.
Due to the asymptotic behavior of the gamma function (see, e.g.,
Eq.~8.328~1.\ in~\cite{Gradshteyn:2007}),
\begin{equation}
  \label{eq:Gamma_asy}
  |\Gamma(x+iy)| \underset{|y|\to\infty}{\sim}
  \sqrt{2\pi} \, e^{-|y| \pi/2} \, |y|^{x-1/2}
  \qquad [x,y \in \mathbb{R}]
  \,,
\end{equation}
the integrands in (\ref{eq:Barnes1}) and~(\ref{eq:Barnes2}) are
exponentially damped for $z \to \pm i\infty$ and the integrals converge.

In the context of loop calculations in quantum field theory, Barnes
contour integrals arise when using Mellin--Barnes (MB) representations by
applying the basic identity
\begin{equation}
  \label{eq:MBid}
  \frac{1}{(A+B)^\lambda} =
  \frac{1}{\Gamma(\lambda)} \MBint{z} \, \frac{B^z}{A^{\lambda+z}} \,
    \Gamma(-z) \, \Gamma(\lambda+z)
  \,,
\end{equation}
with the above-mentioned definition of the integration contour.
In particular, if $\Rep\lambda > 0$, the contour can be chosen as a
straight line in the strip $-\Rep\lambda < \Rep z < 0$. When the MB
integration in~(\ref{eq:MBid}) is interchanged with other integrations over
parameters contained in $A$ or $B$, the convergence of these integrations
will usually be encoded in the arguments of the resulting gamma
functions. The change in the order of integrations is justified if the real
parts of the arguments of all gamma functions remain positive along the
path of the MB integration. Such MB integrals may be analytically continued
to parameter values $\Rep\lambda < 0$, provided that the MB integration
contours are deformed in accordance with the requirements for Barnes
contour integrals described above, separating right and left poles from all
gamma functions in the integrand.

The identity~(\ref{eq:MBid}) is valid unless $\lambda$ is a nonpositive
integer.
Also the ratio $B/A$ must be away from the negative real axis, otherwise
the exponential damping of the gamma functions for $z \to i\infty$ or $z
\to -i\infty$ is compensated by $(B/A)^z \propto e^{\pm i\pi z}$ and the
integral~(\ref{eq:MBid}) might diverge at one of its boundaries.

For $|A| > |B|$, the MB integration contour can be closed to the right at
$\Rep z = +\infty$, and the series of residues at $z=0,1,\ldots$ reproduces
the Taylor expansion of the left-hand side of~(\ref{eq:MBid}) for $|A| >
|B|$. On the other hand, for $|A| < |B|$, the contour may be closed to the
left at $\Rep z = -\infty$, and the series of residues at
$z=-\lambda,-\lambda-1,\ldots$ coincides with the expansion for $|A| <
|B|$. Therefore the MB representation~(\ref{eq:MBid}) provides a way of
separating a sum of two terms $A$ and $B$ without requiring a particular
hierarchy between them.

The MB representation~(\ref{eq:MBid}) was applied
originally~\cite{Usyukina:1975yg} in order to turn massive Feynman
propagators $1/(k^2-m^2+i0)$ into massless ones by separating the mass term
from the squared momentum, using~(\ref{eq:MBid}) with $A=k^2$ and
$B=-m^2+i0$ (see also, e.g.,
\cite{Boos:1990rg,Davydychev:1990jt,Davydychev:1990cq} as early
references).
Later on parametric integrals derived from Feynman diagrams have been
evaluated by applying as many MB representations as necessary to obtain
results in terms of gamma functions under the MB integrals (see, e.g.,
\cite{Greub:1996tg,Greub:2000sy,Asatryan:2001zw,Bieri:2003ue}).
The construction of such MB representations for Feynman integrals with
planar topologies is automated by the Mathematica tool
\texttt{AMBRE}~\cite{Gluza:2007rt,Gluza:2010rn}.

When evaluating the MB integrations, one has to take care of singularities
occurring in certain parameter limits (like $d \to 4$ in dimensional
regularization). These singularities arise when a right pole comes
arbitrarily close to a left pole, pinching the integration contour which
separates them. Two strategies~\cite{Smirnov:1999gc,Tausk:1999vh} were
formulated to systematically extract such singularities analytically. By
taking a finite number of residues, they both end up with integration
contours which are not pinched any more between right and left poles.
These contours can then be chosen as straight lines parallel to the
imaginary axis and the MB integrations may be performed numerically.
The strategy of~\cite{Tausk:1999vh} is automated in the private computer
code~\cite{Anastasiou:2005cb} and by the Mathematica package
\texttt{MB}~\cite{Czakon:2005rk}. More recently also the strategy
of~\cite{Smirnov:1999gc} has been implemented in the computer program
\texttt{MBresolve}~\cite{Smirnov:2009up}.

MB representations are also employed for the asymptotic expansion of
Feynman integrals (see, e.g.,
\cite{Greub:1996tg,Greub:2000sy,Asatryan:2001zw,Bieri:2003ue,%
  Smirnov:2002mg,Friot:2005cu,Jantzen:2011nz,Jantzen:2012mw}). If the MB
integral $\int\!\rd z$ involves the factor~$t^z$, its asymptotic expansion
in the limit $t \to 0$ is obtained by picking up, order by order, the
residues on the right-hand side of the integration contour. Similarly the
asymptotic expansion in the limit $t \to \infty$ results from the residues
on the left-hand side of the contour.
Czakon's code \texttt{MBasymptotics}~\cite{MBTools} performs this
task automatically.
The contributions obtained from asymptotically expanding MB representations
often correspond to those arising in the strategy of expansion by
regions~\cite{Beneke:1997zp,Smirnov:1998vk,Smirnov:1999bza}. This
correspondence is emphasized and used for cross-checks
in~\cite{Jantzen:2011nz,Jantzen:2012mw}. Also combinations of expansion by
regions with further MB representations for evaluating the contributions
have proven fruitful~\cite{Jantzen:2006jv}.

The application of MB representations is explained in many details in the
book~\cite{Smirnov:2006ry}. Computer codes are summarized on the web
site~\cite{MBTools}.

When dealing with multifold MB representations, it is always preferable to
eliminate as many MB integrals as possible using the Barnes lemmas
(\ref{eq:Barnes1}) and~(\ref{eq:Barnes2}) before proceeding to the
extraction of singularities and numerical integrations. So these two lemmas
have become widely used in the application of MB representations. They are
implemented, in addition to previously mentioned tools, in Kosower's
Mathematica package \texttt{Barnes Routines}~\cite{MBTools}.
Let us now turn to the proofs of the two Barnes lemmas.

In his original papers, Barnes focuses on the treatment of hypergeometric
functions. So, naturally, his proofs of the two lemmas employ properties of
such functions.
For proving the first lemma~(\ref{eq:Barnes1}), Barnes~\cite{Barnes:1908}
closes the integration contour to one side, which is only possible for
$\Rep(\alpha_1+\alpha_2+\beta_1+\beta_2) < 1$, and sums the two infinite
series of residues from the corresponding poles. These series are
identified with hypergeometric $\hypF21$ functions of unit argument. After
relating the $\hypF21$ functions to gamma functions, using several times
$\Gamma(x) \, \Gamma(1-x) = \pi/\sin(\pi x)$, and employing a trigonometric
identity, Barnes arrives at the right-hand side of~(\ref{eq:Barnes1}). The
validity of the lemma for general complex values of $\alpha_{1,2}$ and
$\beta_{1,2}$ is argued by analytic continuation.

In his proof of the second lemma~(\ref{eq:Barnes2}),
Barnes~\cite{Barnes:1910} starts with the series representation of a
hypergeometric $\hypF32$ function of unit argument. The number of gamma
functions depending on the summation index is reduced by two using the
first Barnes lemma~(\ref{eq:Barnes1}) from right to left. The series is
then identified with a $\hypF21$ function of unit argument which is written
in terms of gamma functions. The resulting expression involves a contour
integral which can be brought into agreement with the left-hand side
of~(\ref{eq:Barnes2}). This relates the integral from the second lemma to
the $\hypF32$ function, which, for the specific choice of parameters
present in~(\ref{eq:Barnes2}), however, reduces to a $\hypF21$ function of
unit argument and can therefore be expressed through gamma functions,
producing the right-hand side of the second lemma.

These two original proofs by Barnes have been reproduced in several books,
e.g., \cite{Whittaker:1996,Bailey:1935,Slater:2008}. Their understanding
requires knowledge on the series representations and other properties of
hypergeometric functions which appear in intermediary steps.
In the current paper, however, I present new, elegant proofs for the two
Barnes lemmas which only use integral transformations in a straightforward
way, avoiding completely the series summations which the original proofs
employ. My new proofs merely rely on the basic identity of MB
representations~(\ref{eq:MBid}) and on the integral representations of the
gamma function and Euler's beta function. The latter is well known in the
field of loop calculations. Therefore, when introducing MB representations
and the Barnes lemmas to students of quantum field theory, I suggest to use
the proofs presented here rather than the original ones by Barnes.

The new proofs for the two Barnes lemmas are found in
Secs. \ref{sec:Barnes1} and~\ref{sec:Barnes2}, respectively.
Sec.~\ref{sec:lemma3} introduces and proves an additional lemma for a MB
integral involving a phase factor $e^{\pm i\pi z}$, which, apart from its
application to loop calculations, is of particular interest for studying
the convergence behavior of such integrals at their boundaries~$\pm
i\infty$.
Finally conclusions are presented in Sec.~\ref{sec:conc}.

\section{\label{sec:Barnes1}Proof of the first Barnes lemma}

We start with the left-hand side of the first Barnes
lemma~(\ref{eq:Barnes1}),
\begin{equation}
  \label{eq:Barnes1int}
  \MBint{z} \, \Gamma(\alpha_1-z) \, \Gamma(\alpha_2-z) \,
    \Gamma(\beta_1+z) \, \Gamma(\beta_2+z)
  \,.
\end{equation}
Let us assume that
\begin{equation}
  \label{eq:Barnes1cond}
  \Rep(\alpha_i+\beta_j) > 0 \, \forall i,j
\end{equation}
such that the integration contour can be chosen as a straight line parallel
to the imaginary axis with
\begin{equation}
  \label{eq:Barnes1z}
  -\Rep\beta_j < \Rep z < \Rep\alpha_i \, \forall i,j
  \,.
\end{equation}
We insert for $\Gamma(\alpha_1-z)$ and $\Gamma(\beta_1+z)$ the
corresponding integral representations of the gamma function (see, e.g.,
Eq.~8.310~1.\ in~\cite{Gradshteyn:2007}),
\begin{equation}
  \label{eq:Gamma_int}
  \Gamma(\alpha) = \int_0^\infty \! \rd t \, t^{\alpha-1} \, e^{-t}
  \qquad [\Rep\alpha > 0]
  \,,
\end{equation}
and obtain
\begin{align}
  \label{eq:Barnes1Gammarep}
  &\int_0^\infty \! \rd t_1 \, t_1^{\alpha_1-1} \, e^{-t_1}
  \int_0^\infty \! \rd t_2 \, t_2^{\beta_1-1} \, e^{-t_2}
\nonumber \\*
  &\qquad{}\times
  \MBint{z} \left(\frac{t_2}{t_1}\right)^z \,
    \Gamma(\alpha_2-z) \, \Gamma(\beta_2+z)
\end{align}
from~(\ref{eq:Barnes1int}). The change in the order of the integrations is
justified by the condition~(\ref{eq:Barnes1z}) for the contour.
The basic MB identity~(\ref{eq:MBid}) can easily be reformulated by a shift
in the integration variable:
\begin{equation}
  \label{eq:MBidrev}
  \MBint{z} \left(\frac{B}{A}\right)^z \,
    \Gamma(\alpha-z) \, \Gamma(\beta+z)
  = \frac{\Gamma(\alpha+\beta) \, B^\alpha \, A^\beta}{(A+B)^{\alpha+\beta}}
  \,.
\end{equation}
Inserting (\ref{eq:MBidrev}) for the contour integral
in~(\ref{eq:Barnes1Gammarep}) yields
\begin{align}
  \label{eq:Barnes1param}
  \Gamma(\alpha_2+\beta_2)
  \int_0^\infty \! \rd t_1 \int_0^\infty \! \rd t_2 \,
  \frac{t_1^{\alpha_1+\beta_2-1} \, t_2^{\alpha_2+\beta_1-1}}{
    (t_1+t_2)^{\alpha_2+\beta_2}} \,
  e^{-(t_1+t_2)}
  \,.
\end{align}
This double integral factorizes by performing the variable transformation
$t_1 = \eta \, \xi$ and $t_2 = \eta \, (1-\xi)$:
\begin{align}
  \label{eq:Barnes1param2}
  &\Gamma(\alpha_2+\beta_2)
  \int_0^\infty \! \rd\eta \, \eta^{\alpha_1+\beta_1-1} \, e^{-\eta}
\nonumber \\* &\qquad{}\times
  \int_0^1 \! \rd\xi \, \xi^{\alpha_1+\beta_2-1} \, (1-\xi)^{\alpha_2+\beta_1-1}
  \,.
\end{align}
While the $\eta$-integral is given by~(\ref{eq:Gamma_int}), the
$\xi$-integral is a representation of Euler's beta function (see, e.g.,
Eqs. 8.380~1.\ and 8.384~1.\ in~\cite{Gradshteyn:2007}),
\begin{equation}
  \label{eq:Beta_int}
  \int_0^1 \! \rd\xi \, \xi^{\alpha_1-1} \, (1-\xi)^{\alpha_2-1}
  = \frac{\Gamma(\alpha_1) \, \Gamma(\alpha_2)}{\Gamma(\alpha_1+\alpha_2)}
  \qquad [\Rep\alpha_i > 0]
  \,,
\end{equation}
which is an identity often used in loop calculations.
This leads to the result
\begin{equation}
  \label{eq:Barnes1res}
  \frac{\Gamma(\alpha_1+\beta_1) \, \Gamma(\alpha_1+\beta_2) \,
    \Gamma(\alpha_2+\beta_1) \, \Gamma(\alpha_2+\beta_2)}{
    \Gamma(\alpha_1+\alpha_2+\beta_1+\beta_2)}
  \,,
\end{equation}
in agreement with the right-hand side of~(\ref{eq:Barnes1}), and thus
proves the first Barnes lemma.

The transformations in this proof are valid if the
condition~(\ref{eq:Barnes1cond}) is fulfilled, which is when the
integration contour can be chosen as a straight line~(\ref{eq:Barnes1z})
or, equivalently, when the real parts of the arguments of all gamma
functions in the numerator of the result~(\ref{eq:Barnes1res}) are
positive. Through analytic continuation, the validity of the first Barnes
lemma may be extended to general complex values of $\alpha_{1,2}$ and
$\beta_{1,2}$, with the exception of parameter choices where no integration
contour separating right and left poles is found in~(\ref{eq:Barnes1int}),
which corresponds to the right-hand side~(\ref{eq:Barnes1res}) being
ill-defined as well when at least one of the $\alpha_i+\beta_j$ is a
nonpositive integer.

\section{\label{sec:Barnes2}Proof of the second Barnes lemma}

The left-hand side of the second Barnes lemma~(\ref{eq:Barnes2}) reads
\begin{align}
  \label{eq:Barnes2int}
  &\MBint{z} \, \frac{\Gamma(\alpha_1-z) \, \Gamma(\alpha_2-z) \,
      \Gamma(\beta_1+z) \, \Gamma(\beta_2+z)}{
    \Gamma(\alpha_1+\alpha_2+\beta_1+\beta_2+\beta_3+z)}
\nonumber \\*
  &\qquad{}\times
    \Gamma(\beta_3+z)
  \,.
\end{align}
As for the other lemma, we assume that the condition~(\ref{eq:Barnes1cond})
holds for all pairs of parameters such that the integration contour can be
chosen as a straight line in the strip defined by~(\ref{eq:Barnes1z}).
Let us employ the first Barnes lemma~(\ref{eq:Barnes1}) to replace three of
the gamma functions in the integrand:
\begin{align}
  \label{eq:Barnes2Bs}
  &\frac{\Gamma(\beta_2+z) \, \Gamma(\beta_3+z)}{
    \Gamma(\alpha_1+\alpha_2+\beta_1+\beta_2+\beta_3+z)}
\nonumber \\*
  &\quad{} = \frac{1}{\Gamma(\alpha_1+\alpha_2+\beta_1+\beta_2) \,
      \Gamma(\alpha_1+\alpha_2+\beta_1+\beta_3)}
\nonumber \\*
  &\qquad{}\times
    \MBint{s} \, \Gamma(z-s) \, \Gamma(\alpha_1+\alpha_2+\beta_1-s)
\nonumber \\*
  &\qquad{}\times
      \Gamma(\beta_2+s) \, \Gamma(\beta_3+s)
  \,.
\end{align}
Also this second contour integral over~$s$ can be chosen along a straight
line parallel to the imaginary axis with
\begin{equation}
  \label{eq:Barnes2s}
  -\Rep\beta_j < \Rep s < \Rep z < \Rep\alpha_i \, \forall i,j
  \,.
\end{equation}
Then the order of the integrations may safely be interchanged and the
$z$-integral yields
\begin{align}
  \label{eq:Barnes2resz}
  &\MBint{z} \, \Gamma(\alpha_1-z) \, \Gamma(\alpha_2-z) \,
      \Gamma(\beta_1+z) \, \Gamma(-s+z)
\nonumber \\*
  &\quad{} = \frac{\Gamma(\alpha_1+\beta_1) \, \Gamma(\alpha_2+\beta_1) \,
        \Gamma(\alpha_1-s) \, \Gamma(\alpha_2-s)}{
      \Gamma(\alpha_1+\alpha_2+\beta_1-s)}
  \,,
\end{align}
using the first Barnes lemma~(\ref{eq:Barnes1}) again.
Also the final $s$-integral is solved via the first lemma,
\begin{align}
  \label{eq:Barnes2ress}
  &\MBint{s} \, \Gamma(\alpha_1-s) \, \Gamma(\alpha_2-s) \,
      \Gamma(\beta_2+s) \, \Gamma(\beta_3+s)
\nonumber \\*
  &\quad{} = \frac{\Gamma(\alpha_1+\beta_2) \, \Gamma(\alpha_1+\beta_3) \, 
        \Gamma(\alpha_2+\beta_2) \, \Gamma(\alpha_2+\beta_3)}{
      \Gamma(\alpha_1+\alpha_2+\beta_2+\beta_3)}
  \,.
\end{align}
We obtain the result
\begin{align}
  \label{eq:Barnes2res}
  &\frac{\Gamma(\alpha_1+\beta_1) \, \Gamma(\alpha_1+\beta_2) \,
      \Gamma(\alpha_1+\beta_3)}{
    \Gamma(\alpha_1+\alpha_2+\beta_1+\beta_2) \,
      \Gamma(\alpha_1+\alpha_2+\beta_1+\beta_3)}
\nonumber \\*
  &\qquad{}\times
    \frac{\Gamma(\alpha_2+\beta_1) \, \Gamma(\alpha_2+\beta_2) \,
      \Gamma(\alpha_2+\beta_3)}{
    \Gamma(\alpha_1+\alpha_2+\beta_2+\beta_3)}
  \,,
\end{align}
which coincides with the right-hand side of the second Barnes
lemma~(\ref{eq:Barnes2}).
So we have proven the second Barnes lemma simply by applying three times
the first Barnes lemma.

As for the first lemma, the condition~(\ref{eq:Barnes1cond}) can be relaxed
to general complex values of the parameters, as long as none of the
$\alpha_i+\beta_j$ is a nonpositive integer. And, as before, the arguments
of the gamma functions in the numerator of~(\ref{eq:Barnes2res}) reflect
the singularity structure of the integral~(\ref{eq:Barnes2int}).

\section{\label{sec:lemma3}Additional lemma for Mellin--Barnes integrals
  involving a phase factor}

As discussed in the introduction (Sec.~\ref{sec:intro}), the exponential
damping of the gamma function towards $\pm i\infty$~(\ref{eq:Gamma_asy})
usually makes Barnes contour integrals converge well and allows for
numerical integrations of such integrals. This nice feature is potentially
spoiled by phase factors $e^{\pm i\pi z}$ in the integrand. Take, for
instance, the basic MB identity~(\ref{eq:MBid}) with $A=1$ and $B =
-\rho+i0$ ($\rho>0$), thus $B^z = \rho^z \, e^{i\pi z}$. The integrand
still vanishes exponentially for $z \to i\infty$, but at the lower boundary
it behaves as
\begin{align}
  \label{eq:MBid_asy}
  \bigl| \rho^z \, e^{i\pi z} \, \Gamma(-z) \, \Gamma(\lambda+z) \bigr|
  \underset{z \to -i\infty}{\sim}
  2\pi \, \rho^{\Rep z} \, |\!\Imp z|^{\Rep\lambda - 1}
  \,,
\end{align}
so the integral only converges for $\Rep\lambda < 0$. This divergence is
reflected in the left-hand side of~(\ref{eq:MBid}), here
$1/(1-\rho+i0)^\lambda$, which exhibits a singular cancelation for $\rho
\to 1$ if $\Rep\lambda > 0$.

When an ``ordinary'' MB integral (which converges at $z = \pm i\infty$) is
solved in closed form, as for the two Barnes lemmas (\ref{eq:Barnes1})
and~(\ref{eq:Barnes2}), its singularities are encoded in the arguments of
the resulting gamma functions. These singularities originate from right and
left poles pinching the integration contour between them (see
Sec.~\ref{sec:intro}). It is desirable to establish solutions in closed
form also for MB integrals with phase factors $e^{\pm i\pi z}$ where a
potential divergence at $z = \pm i\infty$ is parametrized by the resulting
gamma functions as well.

One such identity is provided by the following \emph{additional lemma}:
\begin{align}
  \label{eq:lemma3}
  &\MBint{z} \, e^{\pm i\pi z} \,
    \frac{\Gamma(\alpha-z) \, \Gamma(\beta_1+z) \, \Gamma(\beta_2+z)}{
      \Gamma(\gamma+z)}
\nonumber \\*
  &\quad{} = e^{\pm i\pi\alpha} \,
    \frac{\Gamma(\alpha+\beta_1) \, \Gamma(\alpha+\beta_2) \,
        \Gamma(\gamma-\alpha-\beta_1-\beta_2)}{
      \Gamma(\gamma-\beta_1) \, \Gamma(\gamma-\beta_2)}
  \,.
\end{align}
While the first two gamma functions in the numerator of the right-hand side
of~(\ref{eq:lemma3}) reflect the singularities originating from right and
left poles, the third gamma function parametrizes the divergence at $z =
\mp i\infty$. In the case of the phase factor $e^{+i\pi z}$, the integrand
in~(\ref{eq:lemma3}) exhibits the asymptotic behavior
\begin{multline}
  \label{eq:lemma3_asy}
  \left| e^{i\pi z} \,
    \frac{\Gamma(\alpha-z) \, \Gamma(\beta_1+z) \, \Gamma(\beta_2+z)}{
      \Gamma(\gamma+z)}
  \right|
\\*
    \underset{z \to -i\infty}{\sim}
    2\pi \, |\!\Imp z|^{\Rep(\alpha+\beta_1+\beta_2-\gamma) - 1}
  \,.
\end{multline}
So the integral only converges at $z = -i\infty$ for
\begin{equation}
  \label{eq:lemma3conv}
  \Rep(\gamma-\alpha-\beta_1-\beta_2) > 0
  \,,
\end{equation}
which matches the argument of the corresponding gam\-ma function
in the right-hand side of~(\ref{eq:lemma3}).

Let us turn to the proof of the additional lemma~(\ref{eq:lemma3}). In
analogy to the previous proofs we assume
\begin{equation}
  \label{eq:lemma3cond}
  \Rep(\alpha+\beta_j) > 0 \, \forall j
\end{equation}
such that the integration contour can be chosen as a straight line parallel
to the imaginary axis with
\begin{equation}
  \label{eq:lemma3z}
  -\Rep\beta_j < \Rep z < \Rep\alpha \, \forall j
  \,.
\end{equation}
Here we also have to require the condition~(\ref{eq:lemma3conv}) for the
convergence of the integral at both boundaries $z=\pm i\infty$.
We start with the left-hand side of~(\ref{eq:lemma3}) and replace two gamma
functions using~(\ref{eq:Beta_int}):
\begin{align}
  \label{eq:lemma3Beta}
  \frac{\Gamma(\beta_2+z)}{\Gamma(\gamma+z)}
  = \frac{1}{\Gamma(\gamma-\beta_2)} \int_0^1 \! \rd\xi \,
    \xi^{\beta_2+z-1} \, (1-\xi)^{\gamma-\beta_2-1}
  \,.
\end{align}
The convergence of the integral~(\ref{eq:lemma3Beta}) follows from
(\ref{eq:lemma3conv}) and~(\ref{eq:lemma3z}). We may safely change the
order of integrations and perform the contour integral first employing the
basic MB identity in the form~(\ref{eq:MBidrev}):
\begin{align}
  \label{eq:lemma3resz}
  &\MBint{z} \,
    \underbrace{e^{\pm i\pi z} \, \xi^z}_{(-\xi \pm i0)^z} \,
    \Gamma(\alpha-z) \, \Gamma(\beta_1+z)
\nonumber \\*
  &\quad{} = \Gamma(\alpha+\beta_1) \,
    \underbrace{(-\xi \pm i0)^\alpha}_{e^{\pm i\pi\alpha} \, \xi^\alpha} \,
    (1-\xi)^{-\alpha-\beta_1}
  \,.
\end{align}
Then the $\xi$-integral is given by~(\ref{eq:Beta_int}):
\begin{align}
  \label{eq:lemma3resxi}
  &\int_0^1 \! \rd\xi \, \xi^{\alpha+\beta_2-1} \,
    (1-\xi)^{\gamma-\alpha-\beta_1-\beta_2-1}
\nonumber \\*
  &\quad{} =
    \frac{\Gamma(\alpha+\beta_2) \, \Gamma(\gamma-\alpha-\beta_1-\beta_2)}{
      \Gamma(\gamma-\beta_1)}
  \,.
\end{align}
By assembling the pieces from (\ref{eq:lemma3Beta}), (\ref{eq:lemma3resz}),
and~(\ref{eq:lemma3resxi}) we obtain the right-hand side
of~(\ref{eq:lemma3}).

The validity of the additional lemma~(\ref{eq:lemma3}) can be extended via
analytic continuation to general complex values of $\alpha$, $\beta_{1,2}$,
and~$\gamma$ by relaxing the condition~(\ref{eq:lemma3cond}), as discussed
for the two previous lemmas. However, the condition~(\ref{eq:lemma3conv})
is crucial for the convergence of the contour integral and cannot be
waived.

\section{\label{sec:conc}Conclusions}

New, elegant proofs have been presented for the two Barnes lemmas which are
often used in loop calculations in order to solve MB integrals. The new
proofs avoid the series summations and representations of hypergeometric
functions which are used in the original proofs by Barnes. They are well
suited for courses in quantum field theory treating MB representations
because they are only based on identities generally known there.

Also an additional lemma has been introduced and proven which parametrizes
the divergence of a MB integral $\int\!\rd z$ when a phase factor $e^{\pm
  i\pi z}$ is present. This new lemma has been used by myself in many
evaluations of Feynman integrals, and other authors have probably solved
this and similar MB integrals as well, e.g.\ by taking a series of residues
and summing them up via a hypergeometric function. But, to my knowledge,
this lemma has not been published in closed form so far.

\begin{acknowledgments}
  The author thanks V.~A.~Smirnov for reading the manuscript and for
  helpful comments.
  This work is supported by the Deutsche For\-schungs\-ge\-mein\-schaft
  Sonder\-forschungs\-bereich/\-Trans\-regio~9 ``Computational Particle
  Physics''.
\end{acknowledgments}

\bibliographystyle{apsrev4-1-bj}
\bibliography{Barnes-Lemmas}

\end{document}